\documentclass[pra,aps,twocolumn]{revtex4}
\usepackage{bm,graphicx,amsmath} \usepackage{bbm}

\newcommand{\abs}[1]{\left|{#1}\right|}
\newcommand{\av}[1]{\left\langle #1 \right\rangle}

\newcommand{\ke}[1]{|#1\rangle}

\newcommand{\var}[2]{\langle #1,#2\rangle}

\newcommand{\al}[1]{^{(#1)}}
\newcommand{\da}{^\dagger}

\newcommand{\pt}[1]{\left( #1 \right)}
\newcommand{\pq}[1]{\left[ #1 \right]}
\newcommand{\pg}[1]{\left\{ #1 \right\}}

\newcommand{\lpq}[1]{\left[ #1 \right.}
\newcommand{\lpg}[1]{\left\{ #1 \right.}

\newcommand{\rpq}[1]{\left. #1 \right]}
\newcommand{\rpg}[1]{\left. #1 \right\}}

\begin{document}

\title{Non-linear optics with two trapped atoms}
\author{Sonia Fern\'{a}ndez-Vidal$^{1,2}$, Stefano Zippilli$^{1,2}$, and Giovanna Morigi$^{1}$}
\affiliation{$^{1}$Departament de F\'{i}sica, Universitat Aut\`{o}noma de Barcelona, E-08193 Bellaterra, Spain\\
$^{2}$ICFO-Institut de Ci\`{e}ncies Fot\`{o}niques, E-08860 Castelldefels, Barcelona, Spain}
\date{\today}

\begin{abstract} We show theoretically that two atomic dipoles in
a resonator constitute a non-linear medium, whose
properties can be controlled through the relative position of the
atoms inside the cavity and the detuning and intensity of the driving laser. We identify the parameter regime where the system operates as a parametric amplifier, based on the cascade emission of the collective dipole of the atoms, and determine the corresponding spectrum of squeezing of the field at the cavity output. This dynamics could be observed as a result of self-organization of laser-cooled atoms in resonators.
\end{abstract}

\maketitle

\section{Introduction}

Quantum light sources are an essential element for implementations
of quantum information processing and secure telecommunication
with quantum optical systems~\cite{qic,Briegel98,Kraus,Braunstein05}.  Experiments have demonstrated several remarkable
milestones, thereby opening promising perspectives for
implementing controlled generation of quantum
light~\cite{Walther04,Kimble03,Kimble04,Rempe02,Kuhn,Lukin04,Kuzmich06,Polzik06,Grangier05,Grangier06,Monroe04,Weinfurter06}.
Besides, these studies touch on the fundamental
question of how macroscopic nonlinear phenomena emerge from the
dynamics of quantum systems~\cite{Savels07}.
A paradigmatic example is the optical parametric amplifier~\cite{WallsMilburn}. This system is usually realized with non-linear crystals in resonators, where the medium response is characterized by the dependence of the macroscopic polarization on the electric field, and where symmetries of the crystal can enhance a certain nonlinear response over others~\cite{Armstrong62,Boyd}. On the other hand, recent theoretical works pointed out that a single atom in a suitable setup can constitute an efficient non-linear optical medium operating in the quantum regime~\cite{Imamoglu97,kimbleParkins,Morigi06,Vitali06}. A question, which naturally emerges from these works, is how these dynamics scale up to a macroscopic non-linear medium, and in particular what is the microscopic building block exhibiting the essential symmetries controlling the order of the medium susceptibility.

\begin{center}\begin{figure}[!th]
    \includegraphics[width=6.5cm]{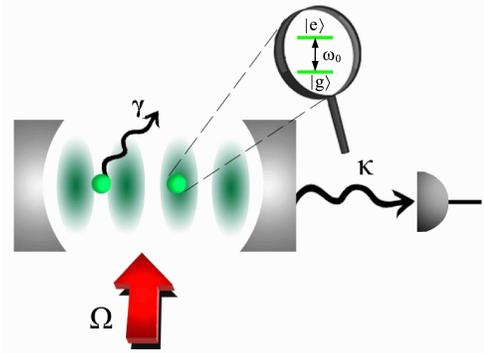}
   \caption{Two atoms are confined inside a high-finesse optical resonator, their
dipoles are driven by a laser and couple to a mode of the cavity. The quantum
state of the field at the cavity output can be controlled by the interatomic
distance inside the resonator and the laser intensity and detuning. A detecting apparatus measures the field at the cavity output.The parameters are defined in Sec.~\protect\ref{Sec:model}. \label{Fig1}}
\end{figure} \end{center}

In this article we study the non-linear response of a medium constituted by two dipoles confined along the axis of an optical resonator, and transversally driven by a laser, in a configuration like the one depicted in Fig.~\ref{Fig1}. At certain interatomic distances the state of the field at the cavity output can exhibit non-classical features. In particular, we show that the system response can be switched from a parametric amplifier to a Kerr medium, just by varying the intensity of the laser field. The validity of our analytical predictions are verified by numerical simulations which take into account the internal dynamics of the atoms and their coupling with the quantized mode of the resonator. The effect of atomic vibrations on the field at the cavity output is estimated using a semiclassical model for the atomic motion. Finally, we discuss the possibility of obtaining such patterns, operating in the quantum regime, as the result of self-organization of laser-cooled atoms in the resonator field~\cite{Domokos02,Asboth05}. 

This article is organized as follows. In Sec.~\ref{Sec:model} the
model is introduced and the basic properties are discussed. In Sec.~\ref{Sec:Heff} the response
of the atomic medium is determined as a function of the atomic
position inside the resonator, when the atoms are driven by a laser. The steady state of cavity and atoms
is determined for the specific parameter regime in
which the system behaves as an optical parametric amplifier. In
Sec.~\ref{Sec:Motion} we consider the effect of the center of mass motion on
the cavity field by means of a semiclassical model. 
In Sec.~\ref{Sec:Conclusions} we
summarize the results and discuss some
outlooks. The appendices provide details of the calculations
presented in Sec.~\ref{Sec:Heff} and Sec~\ref{Sec:Motion}.

\section{ The theoretical model}
\label{Sec:model}

We assume two identical atoms of mass $M$, which are confined inside a
standing-wave cavity, and localized at the position $x_1$ and
$x_2$, respectively, along the cavity axis. We denote by $p_1$ and
$p_2$ the corresponding momenta, and by $H_{\rm mec}$ the
Hamiltonian determining the dynamics of the center of mass in
absence of the coupling with the electromagnetic field, which has
the form \begin{eqnarray}
H_{mec}&=&\frac{p_1^2}{2 M}+\frac{p_2^2}{2 M}+V(x_1,x_2)\\
\end{eqnarray} with $V(x_1,x_2)$ an external potential, which localizes the atoms at their equilibrium positions such that they undergo small vibrations with respect to the cavity-mode wavelength.
The relevant internal degrees of freedom of the
atoms are the ground state $|g\rangle$ and the excited state
$|e\rangle$ of a dipole transition with dipole moment ${\bf d}$, which
is at frequency $\omega_0$. The dipoles are driven by a transverse
laser field at frequency $\omega_L$ and couple to a mode of the
resonator at frequency $\omega_c$ and wave vector $k$, as displayed in Fig.~\ref{Fig1}. 
A detecting apparatus measures the field at the cavity output.

\subsection{Master Equation}

In the
reference frame rotating at the laser frequency the coherent
dynamics of the atoms and cavity mode is described by the
Hamiltonian $H=H_{mec}+H_{at}+H_{cav}+H_{cav-at}+H_{L}$. The terms
\begin{eqnarray} H_{at}&=&-\hbar\Delta
\sum_{j=1,2}\sigma\da_{j}\sigma_{j}\\ H_{cav}&=&-\hbar\delta_{c}a^{\dag}a
\end{eqnarray}
describe the system dynamics in
absence of coupling with the electromagnetic field. Here,
$\Delta=\omega_{L}-\omega_{0}$ and
$\delta_{c}=\omega_{L}-\omega_{c}$ are the detunings of the laser
from the dipole and from the cavity frequency, respectively,
$\sigma_{j}=\left|g\right\rangle_{j}\left\langle e\right|$ the
lowering operator of the atom $j$, $\sigma\da_{j}$ its adjoint,
and $a$ and $a^{\dag}$ are the annihilation and creation operators
of a photon of the cavity mode. The terms
\begin{eqnarray}
\label{Laser}
H_{L}&=&\hbar\Omega\sum_{j=1,2}\left(\sigma\da_{j}+\sigma_{j}\right),\\
H_{at-cav}&=&\hbar \sum_{j=1,2}g(x_j)\left(a^{\dag}\sigma_{j}+ \sigma\da_{j}a\right)
\label{atomfield}
\end{eqnarray} 
describe the interaction of the dipoles with the
cavity and laser fields, respectively, with $\Omega$ the laser
Rabi frequency and $g(x_j)$ the cavity vacuum coupling strength at $x_j$, with
$g(x_j)=g\cos(k x_j)$. In Eq.~(\ref{Laser}) the laser wave vector is orthogonal to the cavity axis.

Coupling to the external environment gives rise to dissipation and decoherence,
which is described by spontaneous emission of the excited state at
rate $\gamma$ and by cavity decay at rate $\kappa$. The dynamics of the 
density matrix $\rho$ of the cavity and atomic degrees of freedom is given by
the master equation
\begin{eqnarray} \label{M:Eq}\frac{\partial}{\partial
t}\rho&=&-\frac{\rm i}{\hbar}\pq{H,\rho}+{\cal
L}_\kappa\rho+{\cal L}_\gamma\rho \\
    &\equiv&{\cal L}\rho\label{L:rho}\end{eqnarray} where
\begin{eqnarray} \label{L:kappa}
    {\cal L}_\kappa\rho=\kappa\pt{2a\rho a\da-a\da a\rho-\rho a\da a}
\end{eqnarray} is the superoperator which describes noise due to
cavity decay, and \begin{eqnarray} \label{L:gamma}{\cal
L}_\gamma\rho=\sum_{j=1,2}\frac{\gamma}{2}\pt{2\sigma_j\tilde{\rho}_j\sigma_j\da-\sigma_j\da\sigma_j\rho-\rho\sigma_j\da\sigma_j}
\end{eqnarray} is the superoperator which describes the quantum
noise due to spontaneous emission. In the superoperator~(\ref{L:gamma}) the term $\tilde{\rho}_j$ accounts for the mechanical effect of the spontaneously emitted photon on the atom in $x_j$, see for instance~\cite{Zippilli05}.

\subsection{Multi-photon processes and atomic patterns}

It is instructive to consider the dynamics in terms of the
collective states of the dipole. We denote by
$|+\rangle$ and $|-\rangle$ the Dicke symmetric and antisymmetric states, respectively, with
$|\pm\rangle=(|eg\rangle\pm|ge\rangle)/\sqrt{2}$, and rewrite the
interaction of the atoms with laser and cavity mode in terms of
the operators \begin{eqnarray}
    S_\pm=(\sigma_1\pm\sigma_2)/\sqrt{2}.
\end{eqnarray} In this representation, the laser-atom interaction, Eq.~(\ref{Laser}),
is rewritten as \begin{equation} H_{L}=\hbar\sqrt{2}\Omega S_+ +{\rm
H.c.} \end{equation} while the atom-cavity interaction term, Eq.~(\ref{atomfield}), can be
decomposed as $H_{at-cav}=H_++H_-$, with
\begin{eqnarray} \label{H:+} H_\pm=\hbar g_\pm(x_1,x_2)\pt{a S_\pm\da+a\da S_\pm}
\end{eqnarray} and \begin{eqnarray}
    g_\pm(x_1,x_2)=\frac{g}{\sqrt{2}}\left(\cos(k x_1)\pm \cos(k x_2)\right).
\end{eqnarray} This decomposition highlights the relevant
cavity-atom dynamics, which depend on the relative atomic
position. The term $H_-$ describes
the coupling of the cavity mode with the Dicke anti-symmetric
state, and it vanishes when the interatomic distance $d=x_2-x_1$ is an integer
multiple of the cavity wavelength $\lambda=2\pi/k$. We denote the corresponding atomic configuration as
"$\lambda$-spaced pattern". The term $H_+$ describes the coupling of the
cavity mode with the Dicke symmetric state and it vanishes when $d$ is
an odd multiple of $\lambda/2$. We denote the corresponding 
atomic configuration as "$\lambda/2$-spaced pattern"
Below we discuss the corresponding dynamics in detail. 

\subsubsection{$\lambda$-spaced pattern.}

We first consider the case in which the interatomic
distance is an integer multiple of $\lambda$. For this configuration, at steady state and for large cooperativities, the atoms are in the ground state and the cavity mode is in a coherent state whose amplitude is determined by the laser intensity~\cite{Zippilli04,Alsing92}. This behaviour can be understood in terms of the coherent buildup of a cavity field, such that its phase is opposite to the driving field. As a result, the atomic dipole is not excited, even if the cavity mode is in a coherent state with a finite number of photons. When two or more atoms are present inside the resonator, this situation can be achieved when the atoms scatter in phase into the cavity modes, i.e., when they are arranged in a $\lambda$-spaced pattern. The coherent scattering processes which two atoms undergo are sketched in Fig.~\ref{fig:ScatteringProcesses}(a) in the Dicke basis, showing that the antisymmetric state $\ke{-}$ remain always decoupled from the coherent dynamics. Here, one identifies the suppression of excitation of the atoms at steady state as due to interference between the excitation path $|gg,n\rangle\to |+,n\rangle$, driven by the laser, and the excitation path $|gg,n+1\rangle\to |+,n\rangle$, driven by the cavity. Figure~\ref{fig:ScatteringProcesses}(a) displays also the other higher-order processes. In particular, we note the processes which lead to the excitation of the state $\ke{ee,n}$ by the absorption of two laser photons, followed by emission of pair of photons into the cavity. These processes are expected to give rise to squeezing of the coherent state of the cavity field. We note that squeezed-coherent radiation has been predicted in the resonance fluorescence of an atomic crystal, at wave vectors such that the Bragg condition of the atomic crystal is equivalent to the $\lambda$-spaced pattern here discussed~\cite{Vogel85}.
Finally, we note that the formation of $\lambda$-patterns of laser-cooled atoms inside of resonators has been predicted as the result of a self-organizing process~\cite{Domokos02,Asboth05}, and features of the field at the cavity output, associated with their formation, have been
measured in~\cite{Chan03,Black03}. Theoretical works have shown that these patterns can be also stable in the strong coupling regime, under the condition, in which atomic excitation is suppressed and the cavity field is in a coherent state~\cite{Zippilli04,Asboth04}.

\subsubsection{$\lambda/2$-spaced pattern.}

\begin{center} \begin{figure}[!th]
\includegraphics[width=8cm]{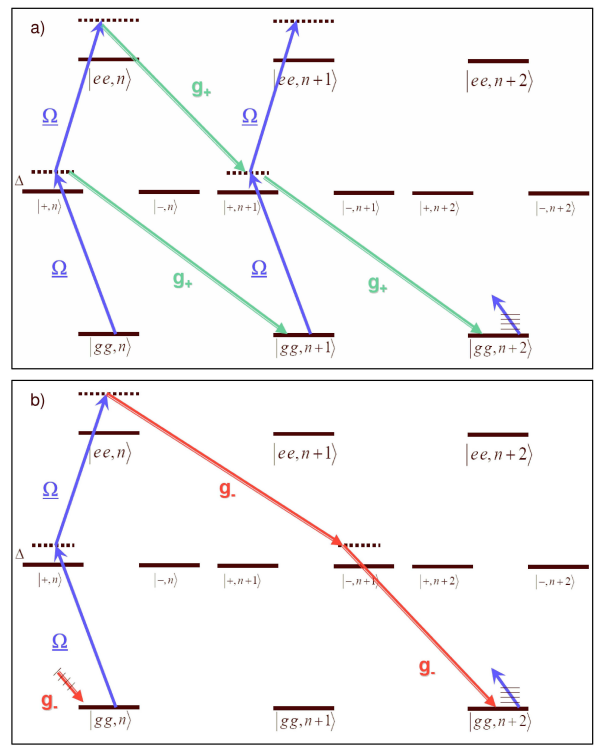}
\caption{Sketch of the
coherent scattering processes between the collective states of two
atomic dipoles driven by a laser and coupled to the cavity mode,
when (a) the interatomic distance $d$ is an integer multiple of
the cavity-mode wavelength $\lambda$, and (b) when $d$ is an odd
multiple of $\lambda/2$. The states $|J,n\rangle$ are the Dicke
states of the two dipoles $|J\rangle$ at $n$ cavity photons, where
$|J\rangle=|gg\rangle,|\pm\rangle~,|ee\rangle$, and
$|\pm\rangle=(|eg\rangle\pm |ge\rangle)/\sqrt{2}$. The arrows
labeled by $\Omega$ ($g_{\pm}$) indicate the transitions driven by
the laser (the cavity mode).} \label{fig:ScatteringProcesses}
\end{figure}\end{center}

We now analyze the case, when the interatomic distance is an odd
multiple of $\lambda/2$, such that $H_+=0$. In this case the
atomic ground state couples via the laser to the Dicke symmetric
state $|+\rangle$, and via the cavity to the antisymmetric state
$|-\rangle$, as depicted in Fig.~\ref{fig:ScatteringProcesses}(b).
Hence, when the laser drives the atoms well below saturation,the cavity is empty~\cite{Zippilli04}. 
In fact, in this limit the two atoms scatter the laser photons with opposite phase into the cavity and the
resulting field vanishes due to destructive interference. 
Figure~\ref{fig:ScatteringProcesses}(b) shows, however, that
the cavity mode can be pumped by higher-order processes, which excite the state $|ee,0\rangle$. In this regime, 
the collective dipole can emit photons in pairs into the cavity mode. These processes are expected to give rise to squeezing of the state of the cavity field. We note that squeezed radiation has been predicted in the resonance fluorescence of an atomic crystal, at wave vectors such that the Bragg condition of the atomic crystal is equivalent to the $\lambda/2$-spaced pattern here discussed~\cite{Vogel85}. In this paper we will investigate the quantum state of the light in presence of a high-finesse cavity when the atoms are
initially in a $\lambda/2$ spaced pattern, and determine the dynamics resulting from the competition between 
coherent processes and noise, such as cavity
decay, spontaneous emission, and atomic vibrations at the equilibrium positions. 

\section{Non-linear response of two trapped atoms}\label{Sec:Heff}

In this section, starting from Eq.~(\ref{M:Eq}) we derive the equation describing the
effective dynamics of the cavity mode in the limit of large atom-laser detuning $|\Delta|\gg g,\Omega,|\delta_c|,\gamma,\kappa$. In this analysis we neglect the effect of atomic motion, and identify the parameter regime in which the system operates as a parametric amplifier. The prediction of the analytical model are compared with the results of a numerical simulation, which evaluate the cavity mode state by solving Eq.~(\ref{M:Eq}).

\subsection{Effective Hamiltonian}

We derive the effective Hamiltonian $H_{\rm eff}$ for the coherent cavity
dynamics at fourth order in the expansion in the small parameters $g/|\Delta|$, $\Omega/|\Delta|$. 
In the Hilbert subspace subtended by the states $\ke{gg,n}$, with $n$ the number of cavity photons, it has the form \begin{eqnarray}\label{Heff}
H_{\rm eff}&=&\pt{\bar\theta-\delta_{c}}a^{\dag}a+\bar\beta\pt{a\da+a}\nonumber\\
& &\bar\chi a^{\dagger}a^{\dagger}aa + \frac{\bar\alpha}{2}\pt{{a\da}^2+{a}^2},
\end{eqnarray} where
\begin{eqnarray}
\bar\theta&=&\frac{g_{+}^{2}
\left(x_{1},x_{2}\right)+g_{-}^{2}\left(x_{1},x_{2}\right)}{\Delta}\\
\bar\beta&=&\frac{\sqrt{2}\Omega}{\Delta} g_{+}\left(x_{1},x_{2}\right) \\
\bar\chi&=&\frac{1}{\Delta^{3}}
\pq{g_{+}^{2}\left(x_{1},x_{2}\right)-g_{-}^{2}\left(x_{1},x_{2}\right)}^2\\
\label{baralpha}
\bar\alpha&=&\frac{2\Omega^{2}}{\Delta^{3}}
\pq{g_{+}^{2}\left(x_{1},x_{2}\right)-g_{-}^{2}\left(x_{1},x_{2}\right)}
\end{eqnarray}
Here, $\bar \theta$ is the a.c.-Stark shift
experienced by the cavity field due to the interaction with the
atoms, the term
$\bar \beta$ comes from the $H_+$ term, Eq.~(\ref{H:+}), and results from the
two-photon transitions coupling the photon states $\ke{n}$ and
$\ke{n\pm 1}$, see Fig.~\ref{fig:ScatteringProcesses}(a). The
amplitude $\bar \alpha$ is the strength of the effective nonlinear pumping of the cavity field
which gives rise to a $\chi^{(2)}$ nonlinearity, typical of a degenerate 
parametric amplifier~\cite{WallsMilburn}. This term is the sum of two contributions, which are
weighted by $g_+$ and $g_-$, respectively, and which represent the
coherent sum of the four-photon processes coupling the states $|gg,n\rangle\to |gg,n\pm 2\rangle$ and depicted in
Fig.~\ref{fig:ScatteringProcesses}(a) and~(b). Finally, the amplitude $\bar\chi$ is the
a.c-Stark shift associated with four-photon processes, where two
cavity photon are virtually absorbed and then emitted along the
transition $|gg,n\rangle\to |ee,n-2\rangle$. This term is present
in both patterns, and gives rise to the $\chi^{(3)}$ nonlinearity
typical of a Kerr medium.

The form of Hamiltonian~(\ref{Heff}) highlights how the two
patterns we considered, $\lambda$- and $\lambda/2$-spaced,
contribute to the various nonlinear processes. We first notice
that in presence of only one atom (when, e.g., $g(x_2)=0$) the
terms $\bar\alpha$ and $\bar\chi$ trivially vanish: these types of
nonlinearities can be clearly generated only when both atoms
couple to the cavity mode. Then, one observes that
the two patterns gives rise to different nonlinear dynamics.
In the $\lambda$-spaced pattern, for instance, all terms in Eq.~(\ref{Heff}) contribute to determine the coherent dynamics
of the cavity mode. While the linear shift $\bar\theta$ can be set to zero by properly choosing the detuning $\delta_c$, on the other hand the linear term scaling with $\bar\beta$ is dominant, and one reasonably expects that it will determine the
cavity steady state.

When the atoms are distributed in a $\lambda/2$-spaced pattern,
the linear drive in Hamiltonian~(\ref{Heff}) vanishes, i.e.,
$\bar\beta=0$, while the only terms which contribute to the coherent dynamics
are at fourth order in the perturbative expansion. Two
possible scenarios can be here identified. (i) When the laser
drive is much weaker than the cavity coupling, $\Omega\ll g$, then $|\bar{\chi}|\gg|\bar{\alpha}|$ and the
dynamics will be basically equivalent to a Kerr medium as in~\cite{Imamoglu97}, whereby in our case the Kerr nonlinearity
emerges from the interaction of the cavity field with the collective dipole of the
atoms. (ii) When the laser drive is much stronger than the cavity
coupling, $\Omega\gg g$, then $|\bar{\chi}|\ll|\bar{\alpha}|$ and the dynamics will be essentially
equivalent to the one in a $\chi^{(2)}$-medium. This is the case
on which we focus in the rest of this article.

\subsection{Realization of a $\chi^{(2)}$ medium}\label{squeezing}

We now consider Hamiltonian~(\ref{Heff}) when the atoms are
localized at the antinodes of the cavity modes in a
$\lambda/2$-spaced pattern, i.e., when $\bar \beta=0$, and when
$\Omega\gg g$, i.e., $|\bar\alpha|\gg|\bar\chi|$. Setting
${\delta}_c=\bar\theta$, the effective coherent dynamics of
the cavity mode is described by Hamiltonian $H_{\rm eff}\approx
H'$, with \begin{equation}\label{h-eff}
H'=\frac{\bar\alpha}{2}\left(a^{2}+ a^{\dag^{2}}\right)
\end{equation} and $\bar\alpha=\alpha$, where now
\begin{eqnarray}\label{alpha}
\alpha=-\frac{4\Omega^{2}g^{2}}{\Delta^{3}} \end{eqnarray} A master equation for the reduced density matrix $\varrho$
of the cavity mode can be derived from Eq.~(\ref{M:Eq}), which
takes the form \begin{eqnarray} \label{M:Eq:1}
\frac{\partial}{\partial t}\varrho
                                  =-\frac{\rm
i}{\hbar}\pq{H',\varrho}+{\cal L}_\kappa\varrho+\tilde{\cal
L}_\gamma\varrho \nonumber\end{eqnarray} where superoperator ${\cal
L}_{\kappa}$ is defined in Eq.~(\ref{L:kappa}), while
\begin{eqnarray} \label{L:gamma1} \tilde{\cal L}_\gamma\varrho
=\frac{\gamma^{\prime}}{2}(2a\varrho
a^{\dagger}-a^{\dagger}a\varrho-\varrho a^{\dagger}a)\end{eqnarray}
describes the damping of the cavity mode via spontaneous emission,
with $\gamma^{\prime}\approx\gamma g^2/\Delta^2$. 

When $\alpha>\kappa+\gamma^{\prime}$,
Eq.~(\ref{M:Eq:1}) predicts that the energy of the cavity mode
increases exponentially as a function of time. Clearly, this exponential
increase is a good approximation only for short times, when the number
of photons inside the cavity mode still warrants the validity of the perturbative
expansion, while for longer times the dynamics will be determined by competition with other processes
which we neglected in the derivation.

When $\kappa+\gamma^{\prime}>\alpha$, a steady state solution
exists, and the corresponding stationary average photon number is
\begin{equation}\label{n0} n_{0}\equiv\left\langle
a^{\dag}a\right\rangle_{\rm
St}=\frac{1}{2}\frac{\alpha^{2}}{\kappa^{\prime 2}-\alpha^{2}},
\end{equation} where $\kappa^{\prime}=\gamma^{\prime}+\kappa$. In
this case, the field quadrature \begin{eqnarray}
    X(t)=a(t){\rm e}^{-{\rm i}\phi}+a\da(t){\rm e}^{{\rm i}\phi}
\end{eqnarray} is squeezed for $\phi=\pi/4$, and its steady-state variance,
$\av{\Delta X_{\rm St}^2}=\av{{X}^2}_{\rm St}-\av{X}^2_{\rm St}$,
takes the form \begin{eqnarray}
        \av{\Delta X_{\rm St}^2}=
    \frac{ \kappa' }{\kappa'+\alpha}.
\end{eqnarray}
Hence, in this case the reduction of the noise
of the quadrature at steady state is such that $\av{\Delta X_{\rm St}^2}>\frac{1}{2}$, since $\kappa^{\prime}>\alpha$.

We now identify parameter regimes in which these
dynamics can be found. Master Equation~(\ref{M:Eq:1}) has
been determined by evaluating the coherent processes up to fourth
order, treating cavity decay at lowest order, and spontaneous
emission at second order in the perturbative expansion. In particular, by deriving the superoperators in
Eq.~(\ref{L:kappa}) and Eq.~(\ref{L:gamma1}) we neglected dissipative scattering processes at higher order in the expansion in $\Omega/|\Delta|, g/|\Delta|$. This is
valid provided that $g^2/|\Delta| > \kappa,\gamma$ and when
 $\alpha\gtrsim \gamma^{\prime}$, which corresponds
 to the condition
\begin{equation}\label{gamma}
 \gamma \lesssim \frac{\Omega^{2}}{\left|\Delta\right|}
\end{equation}
where we used Eq.~(\ref{alpha}). For a dipole
transition with linewidth $\gamma/2\pi=100$~kHz, in a cavity with
$g/2\pi=2.7$MHz, setting $\Omega/2\pi=10$~MHz,
$\left|\Delta\right|/2\pi=100$~MHz. we find
$|\alpha|/2\pi\approx 3$~kHz and a negligible rate of spontaneous
decay. Appreciable squeezing could be observed for a cavity decay
rate of few kHz, which is a demanding experimental condition. 
We will focus on this parameter regime and check numerically the correctness of the predictions of our analytical model.

\subsection{Squeezing spectrum at the cavity
output} \label{Sec:SqueezSpect}

Assuming that the system is in the regime where $\kappa'>\alpha$, we evaluate the
spectrum of squeezing of the field at the cavity
output, namely~\cite{WallsMilburn}
\begin{eqnarray}\label{S:omega} &&S_{\rm out}(\omega)
=2{\rm
Re}\int_0^\infty dt~{\rm e}^{-{\rm i}\omega t}\\
& &\times \left(\av{X_{\rm
out}(t)X_{\rm out}(0)}_{\rm St}-\av{X_{\rm
out}(t)}_{\rm St}\av{X_{\rm
out}(0)}_{\rm
St}\right)\nonumber\end{eqnarray} where the subscript
${\rm St}$ indicates that the averages are performed
over the steady state density matrix. In Eq.~(\ref{S:omega}) $X_{\rm out}(t)$ is
the quadrature of the output field, defined as
\begin{eqnarray}\label{X}
    X_{\rm out}(t)=a_{\rm out}(t){\rm e}^{-{\rm i}\phi}+a_{\rm out}\da(t){\rm e}^{{\rm i}\phi},
\end{eqnarray} with $\phi=\pi/4$ and where
\begin{equation}
\label{a:out}
a_{\rm out}(t)=\sqrt{\kappa}a(t)-a_{\rm in}(t)\end{equation}
and $a_{\rm in}(t)$ is the input noise which is
delta-correlated, $\av{a_{\rm in}(t)a_{\rm in}^{\dagger}(t')}=\delta(t-t')$.
Using the effective model in Eq.~(\ref{M:Eq:1}) we
find an analytical expression of the squeezing spectrum
\begin{eqnarray}\label{sqPO}
S_{\rm out}(\omega)=1-\frac{4\kappa
\alpha}{\pt{\kappa'+\alpha}^2+\omega^2},
\end{eqnarray}
showing that a large reduction of the quadrature fluctuations below the
shot noise limit is achieved at $\omega=0$ when
$\kappa'\approx\alpha$.

\begin{figure}[!th] \begin{center}
\includegraphics[width=8cm]{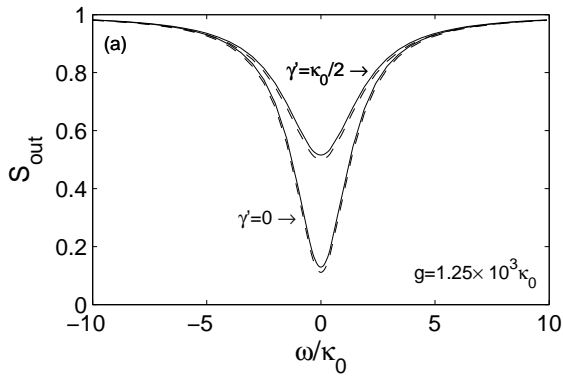}
\caption{Squeezing spectrum of the field at the cavity output, when the atoms are in a $\lambda/2$-pattern. The dashed lines correspond to the spectrum evaluated analytically from Eq.~(\protect\ref{sqPO}), the solid lines to the spectrum  found from the numerical evaluation of the steady state of Eq.~(\protect\ref{M:Eq}), see App.~\ref{app:Spect}. The frequency is in units of $\kappa_0=\kappa$. The parameters are $\Delta=-1.25\times 10^5\kappa_0$, $\Omega=1.25 \times10^4\kappa_0$, $g=1.25\times 10^3\kappa_0$ and $\delta_c=-24\kappa_0$ ($\delta_c$ is chosen so to compensate all a.c.-Stark shifts). For the choice of these parameters, $\alpha=\kappa/2$. The lower  and upper curves have been evaluated for $\gamma=0$ and $\gamma^{\prime}=\kappa_0/2$ ($\gamma=10^4\kappa_0$), respectively.}\label{figS1}\end{center}
\end{figure}

\begin{figure}[!th] \begin{center}
\includegraphics[width=8cm]{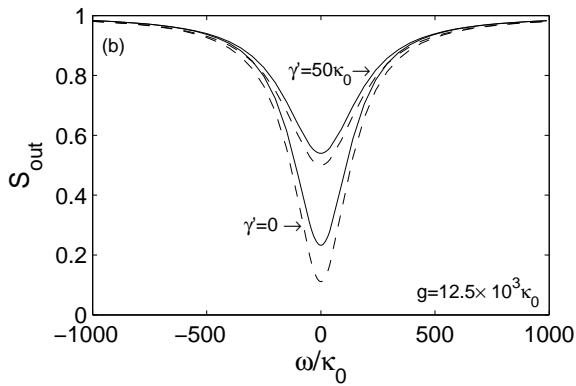}
\caption{Same as in Fig.~\protect\ref{figS1}, where now $\kappa=100\kappa_0$, $g=1.25 \times 10^4\kappa_0$, and $\delta_c=-24\times 10^2\kappa_0$.
For the choice of these parameters, $\alpha=\kappa/2$. The discrepancy between analytical and numerical results is due to the contribution of the Kerr-nonlinearity, which is not accounted for in the analytical model.}\label{figS1:b}\end{center}
\end{figure}

Figure~\ref{figS1} displays the spectrum of squeezing, comparing the
analytical prediction in Eq.~(\ref{sqPO}) with the numerical result obtained
using Eq.~(\ref{M:Eq}), hence including the full internal dynamics
of cavity and atoms, as well as the incoherent processes due to
cavity decay and atomic spontaneous emission at all orders, as discussed in App.~\ref{app:Spect}.
The spectra are evaluated by setting $\alpha=\kappa/2$, and show that for this parameter regime the analytical model provides a good description of the dynamics. We note, as expected, that spontaneous emission tends to decrease the squeezing at the cavity output. Figure~\ref{figS1:b} displays the spectra of squeezing for a larger value of the cavity coupling strentgh. Discrepancies between the analytical and the numerical model arise from the contribution of the Kerr non-linearity in Eq.~(\ref{Heff}), which is not negligible for this parameter regime, since the laser Rabi frequency $\Omega$ and the cavity coupling strength $g$ are of the same order of magnitude.

\begin{figure}[!th] \begin{center}
\includegraphics[width=8cm]{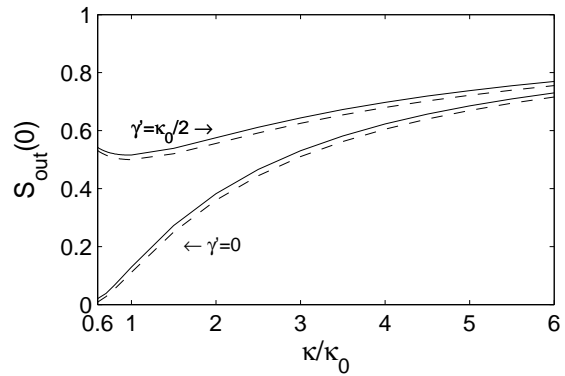}
\caption{Value of the squeezing spectrum $S_{\rm out}(\omega)$ at $\omega=0$, $S_{\rm out}(0)$, as a function of the cavity decay rate $\kappa$ in units of $\kappa_0$.
The dashed lines correspond to the value predicted from Eq.~(\protect\ref{sqPO}), the solid lines to the numerical result found from Eq.~(\protect\ref{M:Eq}). The parameters are $g=1.25\times 10^3\kappa_0$, $\Delta=-1.25\times 10^5\kappa_0$, $\Omega=1.25 \times10^4\kappa_0$, and $\delta_c=-24\kappa_0$. For these parameters $\alpha=\kappa_0/2$. The lower and upper curves have been evaluated for $\gamma=0$ and $\gamma^{\prime}=\kappa_0/2$ ($\gamma=10^4\kappa_0$), respectively.}\label{figS2}\end{center}
\end{figure}

\begin{figure}[!th] \begin{center}
\includegraphics[width=8cm]{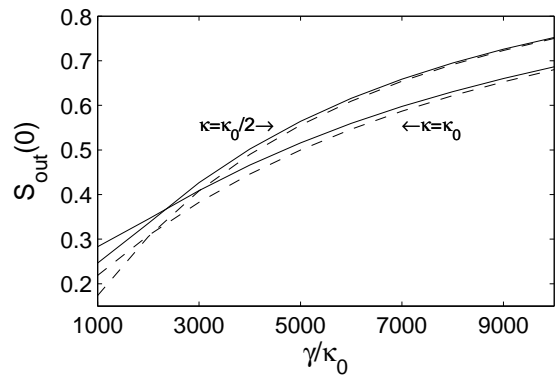}
\caption{$S_{\rm out}(0)$ as a function of the atomic spontaneous emission rate $\gamma$ in units of $\kappa_0$. The spectra are plotted  for two values of the cavity decay rate $\kappa=\kappa_0$ and $\kappa=\kappa_0/2$. The other parameters are as in Fig.~\protect\ref{figS2}. }\label{figS3}\end{center}
\end{figure}

Figure~\ref{figS2} displays the value of the squeezing spectrum at $\omega=0$ as a function of the cavity decay rate $\kappa$.
The spectrum is plotted for $\kappa> \alpha$, when the analytical model described by Eqs.~(\ref{M:Eq:1}) allows for a steady state solution, and it clearly shows that squeezing at the cavity output is very sensitive to variations of $\kappa$. On the other hand, the dependence on the atomic linewidth $\gamma$ is comparatively weak, as one can see from Fig.~\ref{figS3}. The discrepancy between numerical and analytical model at lower values of $\gamma$ is due to the contribution of incoherent scattering processes at higher order, which are accounted for in the numerics and give rise to a very narrow peak at $\omega=0$ in $S(\omega)$. This feature however does not appear for shorter integration times, corresponding to the limit of validity of our perturbative treatment.

\begin{figure}[!th] \begin{center}
\includegraphics[width=8cm]{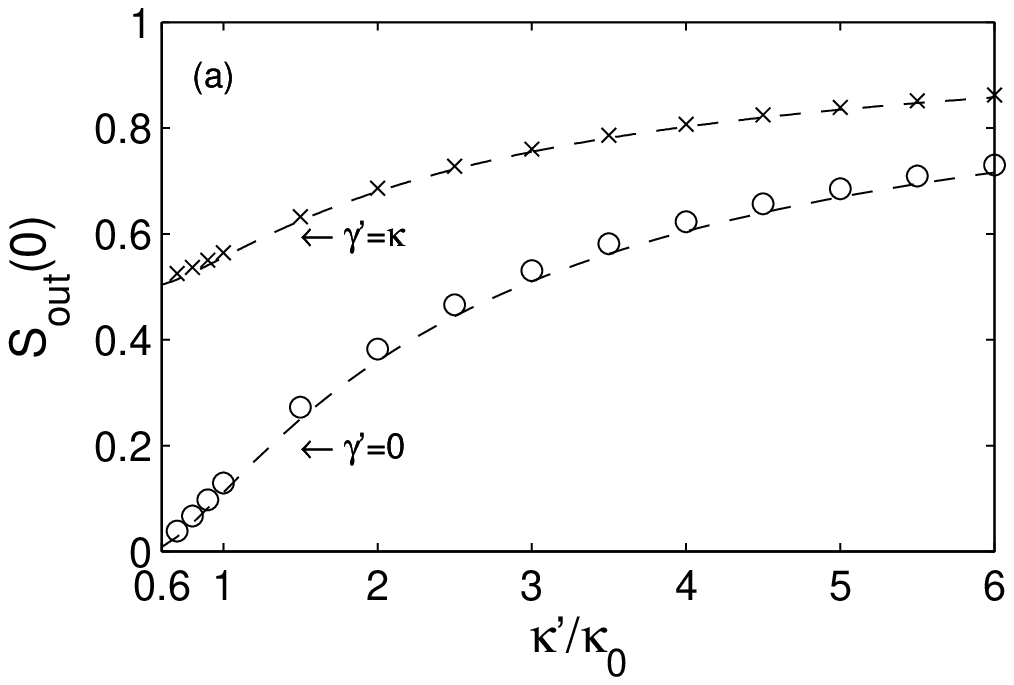}
\includegraphics[width=8cm]{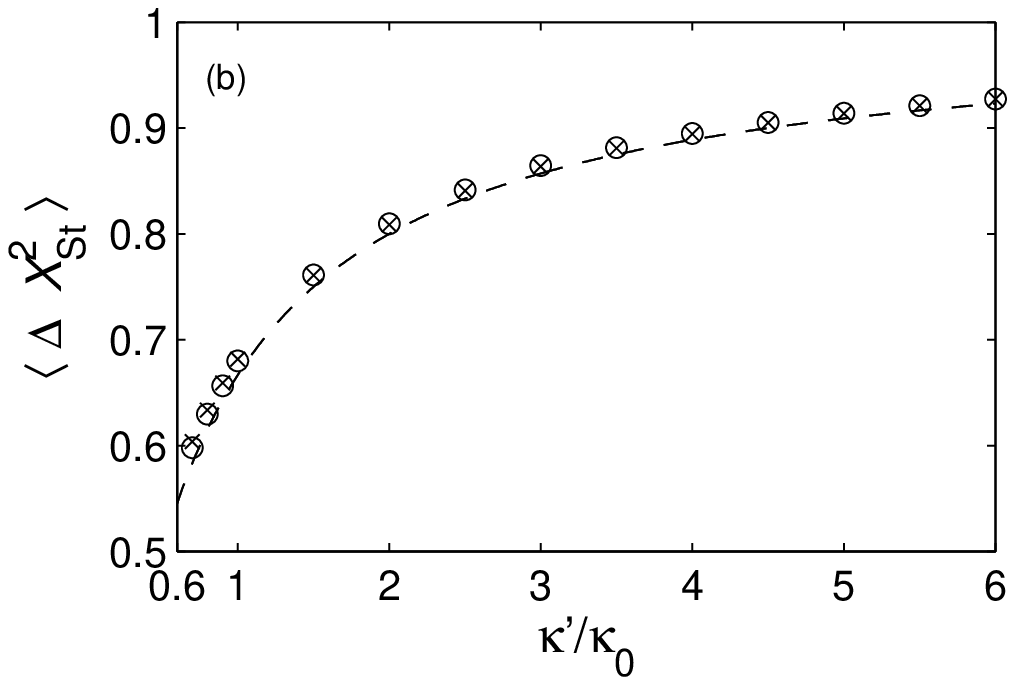}
\caption{(a) $S_{\rm out}(0)$ as a function of the total effective dissipation rate $\kappa'=\kappa+\gamma'$ in unit of $\kappa_0$ and (b) corresponding variance of the squeezed quadrature of the cavity field. The numerical results are displayed for $\kappa=\kappa'$, $\gamma=0$ (circles) and $\kappa=\gamma'=\kappa'/2$ (crosses). The dashed lines are obtained from the analytical model. The other parameters are $g=1.25\times 10^3\kappa_0$, $\delta_c=-24\kappa_0$, $\Delta=-1.25\times 10^5\kappa_0$, $\Omega=1.25 \times10^4\kappa_0$, and $\alpha=\kappa_0/2$.
}\label{figS4}\end{center}
\end{figure}

Figures~\ref{figS4}(a) and~(b) display the spectrum of squeezing at $\omega=0$ and the corresponding variance of the maximally squeezed quadrature of the cavity field as a functions of $\kappa'=\kappa+\gamma'$. In Fig.~\ref{figS4}(a) the upper curves are obtained for $\kappa=\gamma'=\kappa'/2$, the lower curves correspond to $\gamma'=0$, $\kappa'=\kappa$. Figure~\ref{figS4}(b) shows that the variance of the quadrature is the same both for $\gamma'=0$ and $\gamma'=\kappa$, showing that spontaneous emission in this regime only dissipates the squeezed field along other channels, as predicted from the analytical model of Eq.~(\ref{M:Eq:1}).
\section{Effect of the atomic motion} \label{Sec:Motion}

So far we have studied the dynamics of the cavity mode neglecting the atomic
kinetic energy on the cavity-mode dynamics. In this section we
study the effect of fluctuations in the atomic positions,
when the system operates as an optical parameteric amplifier.
We assume that the atoms are confined by an external potential, which
localize them at the antinodes of the cavity standing wave in a
$\lambda/2$-spaced pattern, in the regime in which the mechanical
effects of the cavity field on the atomic motion can be neglected. This situation could be realized experimentally with the technology developed for instance in~\cite{Guthohrlein,Rauschenbeutel,Kuhn05,Chapman07}.

Denoting by $\bar x_j$ the atomic equilibrium positions, and 
by $q_j=x_j-\bar x_j$ the displacements, we write the external potential for small vibrations
as
 \begin{eqnarray}
    V(x_1,x_2)=\frac{1}{2}M\nu^2\pt{q_1^2+q_2^2}.
\end{eqnarray} where $\nu$ is the trapping frequency. The
Heisenberg-Langevin equation of motion for the atomic displacement
$q_j$ is given by~\cite{HelmutFP} \begin{equation} \label{Eq:motion}
\ddot{q}_j=-\nu^2q_j-\frac{F\al{j}}{M}+\xi(t) \end{equation} where
$\xi(t)$ is the quantum Langevin force, associated with the
spontaneous emission and the cavity decay
processes, and \begin{equation} \label{force}
{F}\al{j}=\frac{\partial}{\partial x_j}H_{at-cav} \end{equation}
is the mechanical force operator arising from the spatial gradient
of the atom-cavity interaction over the atomic wave packet. These
equations have to be solved together with the Heisenberg-Langevin
equations for the field, which depend on the atomic motion through
the functions $\cos kx_j$. We assume that the atoms are well
localized at the antinodes of the cavity mode, namely that $\delta
q=\sqrt{\av{q_j^2}}\ll \lambda$, and make a perturbative expansion
in the small parameter $k\delta q$. At second order, the equations
for the fields read \begin{eqnarray}\label{Eq:a}
\dot{a}(t)&=&-{\rm i} \alpha a^{\dag}(t)-\pq{\kappa'+{\rm i}\pt{\theta-\delta_c}} a(t) + \eta\left(t\right)\\
&&
+{\rm i}\frac{k^2}{2}\pq{\pt{q_1^2+q_2^2}\pt{\alpha a\da(t)+\theta a(t)}+\pt{q_1^2-q_2^2}\beta}\nonumber\\
\label{Eq:ada}
\dot{a\da}(t)&=&{\rm i} \alpha a(t)-\pq{\kappa'-{\rm i}\pt{\theta-\delta_c}} a\da(t) + \eta\left(t\right)\\
&&-{\rm i}\frac{k^2}{2}\pq{\pt{q_1^2+q_2^2}\pt{\alpha a(t)+\theta
a\da(t)}+\pt{q_1^2-q_2^2}\beta}
\nonumber
\end{eqnarray}
with $\alpha$ defined in Eqs.~(\ref{alpha}),
$\beta=g\Omega/\Delta$, $\theta=2g^{2}\Delta$, and
$\eta(t)$ is the quantum Langevin term,
$\eta(t)=\sqrt{2\kappa}a_{\rm in}(t)+\sqrt{2\gamma'}a^{\rm at}_{\rm in}(t)$.
Here $a^{\rm at}_{\rm in}(t)$ is the input noise term associated with atomic spontaneous emission, which satisfy the relation $\av{a^{\rm at}_{\rm in}(t){a^{\rm at}_{\rm in}}^{\dagger}(t')}=\delta(t-t')$.

\begin{figure}[!th] \begin{center}
\includegraphics[width=8cm]{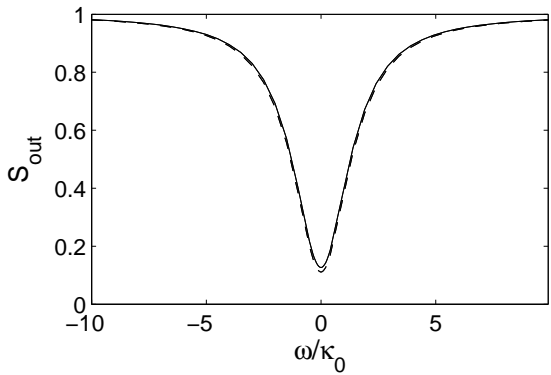} \caption{Spectrum of
squeezing of the field at the cavity output as a function of
$\omega$ in units of $\kappa_0$, for the same parameters as in
Fig.~\ref{figS1} and $\gamma=0$. The solid curve corresponds to
the spectrum of Eq.~(\ref{aa}) for $k\bar q=0.3$. The dashed line
corresponds to the spectrum of Eq.~(\ref{sqPO}) when atomic vibrations
are neglected.}\label{figS5}\end{center} \end{figure}

Even when the atoms are well localized around the antinodes of the
cavity mode, the systematic solution of these coupled equations is
rather complex. Here, we
assume that the external potential provides a steep confinement,
such that the effect of the coupling with the cavity mode can be
neglected in Eq.~(\ref{Eq:motion}). In this limit the solution of
Eq.~(\ref{Eq:motion}) reads  \begin{eqnarray}\label{q(t)}
    q_j(t)\simeq q_j^{(0)}\cos(\nu t+\phi_j),
\end{eqnarray} where $q_j^{(0)}$ and
$\phi_j$ are determined by the initial conditions. When the
trap frequency is much larger than the effective rates which
determine the evolution of the field, $\nu\gg\alpha,\kappa^{\prime}$, we
can derive a secular equation for the cavity field by averaging
the equations for the cavity variables over a period
$T=2\pi/\nu$~\cite{Blumel}. We insert Eq.~(\ref{q(t)}) into the
equations for the field variables,
Eq.~(\ref{Eq:a})-(\ref{Eq:ada}), and integrate them over the
period $T$. With this procedure we find equations for the
operators $\tilde a(t)$, $\tilde \eta(t)$, defined as
\begin{eqnarray}
    \tilde a(t)=\frac{1}{T}\int_t^{t+T}d\tau a(\tau),~~
    \tilde \eta(t)=\frac{1}{T}\int_t^{t+T}d\tau
    \eta(\tau)
\end{eqnarray} Here, the new noise operators satisfy the equation
$\av{\tilde\eta(t) \tilde\eta\da(t')}\simeq 2\kappa'\delta(t-t')$,
where the $\delta$-like correlation is to be interpreted for the
coarse-grained time scale. The corresponding Heisenberg-Langevin
equations read \begin{eqnarray}\label{tilde a}
    \dot{\tilde a}(t)&=& -{\rm i}\tilde\alpha\tilde a\da(t)-\pq{\kappa+{\rm i}\pt{\tilde\theta-\delta_c}}\tilde a(t)+\tilde\eta(t)\\
    \dot{\tilde a}(t)\da&=& {\rm i}\tilde\alpha\tilde a(t)-\pq{\kappa-{\rm i}\pt{\tilde\theta-\delta_c}}\tilde a\da(t)+\tilde\eta\da(t)
    \label{tilde ad}
\end{eqnarray} while their derivation is discussed in
App.~\ref{timeaverage}. Here,
 \begin{eqnarray}
    \tilde \alpha&=&\alpha(1-k^2\bar q^2/2)\nonumber\\
    \tilde \theta&=&\theta(1-k^2\bar q^2/2).\label{theta}
\end{eqnarray} and we have assumed that the oscillation amplitudes
of the two atoms are equal, $q_1^{(0)}= q_2^{(0)}=\bar q$. The
motion-induced a.c.-Stark shift can be compensated by properly
tuning the laser frequency, $\tilde{\delta}_c=\theta(1-k^2\bar
q^2/2)$, and Eqs.~(\ref{tilde a})-(\ref{tilde ad}) become
\begin{eqnarray}\label{aa}
    \dot{\tilde a}(t)&=& -{\rm i}\tilde\alpha\tilde a\da(t)-\kappa'\tilde a(t)+\tilde\eta(t)\nonumber\\
    \dot{\tilde a}(t)&=& {\rm i}\tilde\alpha\tilde a(t)-\kappa'\tilde a\da(t)+\tilde\eta\da(t).
\end{eqnarray} Correspondingly, at lowest order in $k\bar q_j$ the
spectrum of squeezing is \begin{eqnarray} \label{sqPO:motion}
    S_{\rm out}(\omega)&=&1-\frac{4\kappa \alpha}{\pt{\kappa'+\alpha}^2+\omega^2}\nonumber\\
&&\times\pq{1+\frac{\pt{\alpha^2-{\kappa'}^2-\omega^2}}{\pt{\kappa'+\alpha}^2+\omega^2}
\frac{k^2\bar q^2}{2} } \end{eqnarray} where the term proportional
to $k^2\bar q^2$ is the correction to Eq.~(\ref{sqPO}) due to
small vibrations of the atoms at the equilibrium positions. Small
fluctuations hence reduce the bandwidth of frequencies where the
light is squeezed. The corresponding spectrum, Eq.~(\ref{sqPO:motion}),
is displayed in Fig.~\ref{figS5} for $k\bar q=0.3$ and compared to the one of Eq.~(\ref{sqPO}),
where atomic motion is neglected, showing that the
modification of the spectrum of squeezing due to the motion is very small.

\section{Conclusion} \label{Sec:Conclusions}

We have studied the dynamics and steady state of a medium composed
by two atomic dipoles confined inside a resonator in an ordered
structure. Depending on the relative position of the atoms inside
the cavity mode, the linear response can be suppressed, and by
tuning the intensity of the laser the system can operate as Kerr
medium or as optical parametric amplifier, whereby the nonlinear
response emerges from the collective excitations of the atomic
dipoles. We have studied in detail the case in which the system
operates as an optical parametric amplifier, and investigated the
squeezing of the field at the cavity output considering the
effects of atomic vibrations, when the atoms are confined inside the resonator at the
equilibrium positions of a steep external potential, in a situation which
can be experimentally realized for instance in~\cite{Guthohrlein,Rauschenbeutel,Kuhn05,Chapman07}. 

A natural question, emerging from recent studies on selforganization
of laser-cooled atoms in resonators~\cite{Domokos02,Asboth05,Chan03,Black03}, is whether
in absence of an external potential trapping the atoms, the $\lambda/2$-spaced pattern can be
sustained by the mechanical forces of the potential generated by
the scattered field. In~\cite{Asboth04} a semiclassical and numerical analysis showed that this configuration is expected to be
stable for choices of the parameters, which are consistent with the operational regime in which squeezed light can be observed. In this case, one would hence have a selforganized pattern, which sustains and is sustained by non-classical light. 

The results of this work provide an example of how non-linearities
emerge from the microscopic dynamics of few simple quantum
systems. In this respect, two atoms in a resonator can be
considered the most basic realization of a non-linear crystal,
with however limited efficiencies. We conjecture that by scaling
up the number of atoms collective effects can enhance the
nonlinear properties, thus improving the system response. 
Another interesting question is how the system dynamics are modified when
the quantum nature of the atomic motion is
relevant~\cite{Maschler06,Larson07}, and in particular how the
correlation functions of the output field are affected by the
quantum properties of the medium. This study requires an
analysis of the spectrum of resonance fluorescence as
in~\cite{Zippilli07}, which systematically accounts for the
quantum state of the atomic motion, and it will be object of
future investigations.

\acknowledgments The authors are grateful to J\"urgen Eschner,
Helmut Ritsch, Jonas Larson, Maciej Lewenstein, and Roberta
Zambrini, for stimulating discussions and helpful comments.
Support by the European Commission (EMALI, MRTN-CT-2006-035369;
SCALA, Contract No.\ 015714), by the Spanish Ministerio de
Educaci\'on y Ciencia (Consolider Ingenio 2010 QOIT, CSD2006-00019; QLIQS, FIS2005-08257;
Ramon-y-Cajal individual fellowship) are acknowledged.

\begin{appendix}

\section{Evaluation of the Squeezing spectrum}
\label{app:Spect}

Using Eq~(\ref{a:out}), we rewrite the squeezing spectrum in Eq.~(\ref{S:omega}) as
\begin{eqnarray}\label{SpSq}
&&  S_{\rm out}(\omega)=1+4\kappa{\rm Re}\int_0^\infty dt {\rm e}^{-{\rm i}\omega t}\lpq{\var{a(t)}{a(0)}_{\rm St}{\rm e}^{-2{\rm i}\pi/4}  }\nonumber\\
&&  \rpq{
    +\var{a\da(0)}{a\da(t)}_{\rm St}{\rm e}^{2{\rm i}\pi/4}+\var{a\da(t)}{{a(0)}}_{\rm St}+\var{a\da(0)}{a(t)}_{\rm St}
    }.\nonumber\\
\end{eqnarray}
where $$\var{a\da(0)}{a\da(t)}_{\rm St}=\langle a\da(0)a\da(t)\rangle_{\rm St}-\langle a\da(0)\rangle_{\rm St}\langle a\da(t)\rangle_{\rm St}.$$
Equation~(\ref{SpSq}) can be expressed in terms of averages performed over a density matrix by means of the relation
$\av{A(t)A(0)}_{\rm St}={\rm Tr}\pg{A{\rm e}^{{\cal L}t}A\rho_{\rm St}}$ and $\av{A(0)A(t)}_{\rm St}={\rm Tr}\pg{A{\rm e}^{{\cal L}t}\rho_{\rm St}A}$, where $A$ is a generic operator, $\cal L$ is the Liouvillian defined in Eq.~(\ref{L:rho}) setting $H_{mec}=0$, and $\rho_{\rm St}$ is the steady state density matrix satisfying the relation ${\cal L}\rho_{\rm St}=0$. Therefore the spectrum of squeezing can be rewritten  as
\begin{eqnarray}\label{SpSq2}
&&  S\al{\theta}_{\rm out}(\omega)=
1-4\kappa{\rm Re}\left[\pi\delta\left(\omega\right){\rm Tr}\pg{ X\al{\theta}\rho_{\rm St}}^{2} \right.\nonumber\\
&&+\left.{\rm Tr}\pg{      X\al{\theta}\pt{{\cal L}-{\rm i}\omega}^{-1}\pt{a\rho_{\rm St}{\rm e}^{-{\rm i}\theta}+\rho_{\rm St}a\da{\rm e}^{{\rm i}\theta}}} \right]
\end{eqnarray}
The numerical results in Sec.~\ref{Sec:SqueezSpect} are based on the evaluation of the spectrum of squeezing, as calculated from Eq.~(\ref{SpSq2}) using the Liouvillian of Eq.(~\ref{M:Eq}).

\section{Derivation of the secular equations for fast vibrating atoms}
\label{timeaverage}

\noindent After inserting Eq.~(\ref{q(t)}) into the Eqs. Eqs.~(\ref{Eq:a})-(\ref{Eq:ada}), we obtain
\begin{eqnarray}\label{Eq:dotA}
    \dot A(t)&=&MA(t)+N(t)\nonumber\\
    &&+\sum_{j=1,2}k^2\bar q_j^2\cos^2(\nu t+\phi_j)\pq{VA(t)+(-1)^jB}\nonumber\\
\end{eqnarray}
where
\begin{eqnarray}
    A(t)=\pt{\begin{array}{c} a(t) \\ a\da(t)\end{array}},
\end{eqnarray}
\begin{eqnarray}
    M=\pt{\begin{array}{cc}
    -\kappa'-{\rm i}(\theta-\delta_c) & -{\rm i}\alpha \\
    {\rm i}\alpha &   -\kappa'+{\rm i}(\theta-\delta_c)
    \end{array}},
\end{eqnarray}
\begin{eqnarray}
   N(t)=\pt{\begin{array}{c} \eta(t) \\ \eta\da(t)\end{array}},
\end{eqnarray}
\begin{eqnarray}
    V
    =\pt{\begin{array}{cc}
    {\rm i}\theta/2 & {\rm i}\alpha/2 \\
    -{\rm i}\alpha/2 &  -{\rm i}\theta/2
    \end{array}},
\end{eqnarray}
and
\begin{eqnarray}
   B= \pt{\begin{array}{c} -{\rm i}\beta \\ {\rm i}\beta\end{array}}.
\end{eqnarray}
We indicate with
\begin{eqnarray}
     \tilde{f}(t)=\frac{1}{T}\int_t^{t+T} d\tau f(\tau)
\end{eqnarray}
the time average of a variable $f(t)$ over a period of oscillation $T=2\pi/\nu$ of the atomic motion. Since $$\frac{\partial}{\partial t} 
\tilde{f}(t)=\frac{1}{T}\pq{f(t+T)-f(t)}=\widetilde{\frac{\partial f}{\partial t}},$$
we find
\begin{eqnarray}\label{Eq:dotavA}
    \frac{\partial}{\partial t} \tilde{A}(t)
    &=&M \tilde{A}(t)+\tilde{N}(t)+k^2\bar q^2 V\tilde{A}(t)\nonumber\\
    &&+k^2\bar q^2 \frac{V}{2}\sum_{j=1,2}\frac{1}{T}\int_t^{t+T} d\tau{\cos(2\nu \tau+2\phi_j)A(\tau)}\nonumber\\
    \end{eqnarray}
where we have used the relation $\cos^2(y)=\frac{1}{2}\pq{1+\cos(2y)}$ and we have assumed that the two
atoms have the same energy, such that $\bar q_1^2=\bar q_2^2=\bar
q^2$. We now identify the conditions under which we can neglect the second line of Eq.~(\ref{Eq:dotavA}).
Integrating by part the second line of Eq.~(\ref{Eq:dotavA}) an using Eqs.~(\ref{Eq:dotA}) and (\ref{Eq:dotavA}) we obtain
\begin{eqnarray}\label{Eq:dotavAb}
    \frac{\partial}{\partial t} \tilde{A}(t)
        &=&M \tilde{A}(t)+\tilde{N}(t)+k^2\bar q^2 V\tilde{A}(t)\nonumber\\
    &&+k^2\bar q^2C(t)+k^4\bar q^4 D(t)+k^4\bar q^4 E
\end{eqnarray}
where
\begin{eqnarray}
    C(t)&=&\frac{V}{4\nu}\sum_j\lpg{ \sin(2\nu t+2\phi_j)\pq{M \tilde{A}(t)+\tilde{N}(t)} }\nonumber\\
    &&\rpg{- \frac{1}{ T}\int_t^{t+T}d\tau\sin(2\nu \tau+2\phi_j)\pq{M {A}(\tau)+{N}(\tau)} },\nonumber\\
    D(t)&=&\frac{V^2}{8\nu T}\sum_{jj'}\int_t^{t+T} d\tau \nonumber\\
    &&\times  \lpq{\sin(2\nu t+2\phi_j)}\cos(2\nu \tau+2\phi_{j'}) \nonumber\\
    &&\rpq{- 2 \sin(2\nu \tau+2\phi_j)\cos^2(\nu\tau+\phi_{j'})}
    A(\tau), \nonumber\\
    E&=&\frac{VB}{8\nu}\sin[2(\phi_2-\phi_1)].
\end{eqnarray}
The terms $k^2\bar q^2C(t)+k^4\bar q^4 D(t)$ are negligible with respect to $k^2\bar q^2 V$ when $\abs{\theta}\kappa'/8\nu\ll\abs{\alpha}/2$ and $\abs{\theta(\theta-\delta_c)}/8\nu\ll\abs{\alpha}/2$ which reduce to 
 \begin{eqnarray}\label{condition1}
	\nu&\gg& \frac{g^2}{\abs{\Delta}},\\
	\nu&\gg& \frac{k^2\bar q^2}{8}\abs{\frac{g^2\Delta}{\Omega^2}}
	\label{condition1b}
\end{eqnarray}
when  $|\alpha|$ and $\kappa'$ are of the same order of magnitude and $\delta_c=\theta(1-k^2\bar q^2/2)$, see Eq.~(\ref{theta}). 
The term $k^4\bar q^4 E$ in Eq.~(\ref{Eq:dotavAb}) can be neglected when $k^4\bar q^4\abs{\theta\beta}/16\nu\ll k^2\bar q^2\abs{\alpha}/2$,
that is
\begin{eqnarray}\label{condition2} 
\nu &\gg&\frac{k^2\bar q^2 }{16}\abs{\frac{g\Delta}{\Omega}}. 
\end{eqnarray}
When conditions~(\ref{condition1})-(\ref{condition2}) are satisfied we approximate Eq.~(\ref{Eq:dotavA}) with 
\begin{flushleft}

\end{flushleft}
 \begin{eqnarray}
    \frac{\partial}{\partial t} \tilde{A}(t)
    &=&M \tilde{A}(t)+\tilde{N}(t)+k^2\bar q^2V\tilde{A}(t).\nonumber\\
    \end{eqnarray}
which then leads to Eqs.~(\ref{tilde a}) and~(\ref{tilde ad}).
Finally we show that the averaged noise operators, $\tilde
\eta(t)$ and $\tilde \eta\da(t)$, which appear in the term
$\tilde{N}(t)$, are delta correlated. The only non-vanishing
correlation function is \begin{eqnarray}
&&\av{\tilde{\eta}(t)\tilde\eta\da(t')}=\frac{2\kappa'}{T^2}\int_t^{t+T}d\tau\int_{t'}^{t'+T}d\tau'\delta\pt{\tau-\tau'}\nonumber\\
&&=\lpg{\begin{array}{l}
\frac{2\kappa'}{T^2}\pt{t'+T-t}\ {\rm for }\ t'<t<t'+T\\
\frac{2\kappa'}{T^2}\pt{t+T-t'}\ {\rm for }\ t'-T<t<t'\\
0 \ \ \ \ \ {\rm for }\ t>t'+T \ {\rm or}\ t<t'-T
\end{array}}\nonumber\\
\end{eqnarray} which is not zero only if the two integration
intervals $[t,t+T]$ and $[t',t'+T]$ have finite overlap. Therefore
if ${f}(t)$ varies slowly over the time $T$, so that
$\tilde{f}(t)\simeq  f(t) $, then one has \begin{eqnarray}
\int_{-\infty}^\infty dt
f(t)\av{\tilde{\eta}(t)\tilde\eta\da(t')}\simeq 2\kappa' f(t'),
\end{eqnarray} that is $
    \av{\tilde{\eta}(t)\tilde\eta\da(t')}\simeq 2\kappa'\delta(t-t')$. \end{appendix}

\end{document}